\documentclass[aps,prl,superscriptaddress,twocolumn,floatfix,showpacs,a4paper]{revtex4}

\usepackage{graphicx,graphics,epsfig}   % Include figure files
\usepackage{dcolumn}    % Align table columns on decimal point
\usepackage{bm}         % bold math
\usepackage{amsmath}    % need for subequations
\usepackage{verbatim}   % useful for program listings
\usepackage{color}      % use if color is used in text
\usepackage{subfigure}  % use for side-by-side figures
\usepackage{times,natbib}
\usepackage{amsmath,amsfonts,amssymb,graphics,graphicx,epsfig,color,times,natbib}

\begin{document}
\title{Einstein-Podolsky-Rosen Steering in Quantum Phase Transition}

\author{Chunfeng Wu}
\affiliation{Centre for Quantum Technologies, National University of
Singapore, 3 Science Drive 2, Singapore 117543}

\author{Jing-Ling Chen}
 \email{chenjl@nankai.edu.cn}
\affiliation{Centre for Quantum Technologies, National
University of Singapore, 3 Science Drive 2, Singapore 117543}
\affiliation{Theoretical Physics Division, Chern Institute of
Mathematics, Nankai University, Tianjin 300071, People's Republic of
China}

\author{Dong-Ling Deng}
\affiliation{Department of Physics and MCTP, University of Michigan,
Ann Arbor, Michigan 48109, USA}

\author{Hong-Yi Su}
\affiliation{Centre for Quantum Technologies, National
University of Singapore, 3 Science Drive 2, Singapore 117543}
\affiliation{Theoretical Physics Division, Chern Institute of
Mathematics, Nankai University, Tianjin 300071, People's Republic of
China}

\author{X. X. Yi}
\affiliation{Centre for Quantum Technologies, National
University of Singapore, 3 Science Drive 2, Singapore 117543}
\affiliation{School of Physics and Optoelectronic Technology, Dalian University of Technology, Dalian 116024, China}

\author{C. H. Oh}
\email{phyohch@nus.edu.sg}
\affiliation{Centre for Quantum Technologies, National University of
Singapore, 3 Science Drive 2, Singapore 117543}
\affiliation{Department of Physics, National University of
Singapore, 2 Science Drive 3, Singapore 117542}

\date{\today}

\begin{abstract}
We investigate the Einstein-Podolsky-Rosen (EPR) steering and its criticality in quantum phase transition.
It is found that the EPR steerability function of the ground state of XY spin chain exhibits nonanalytic feature in the
vicinity of a quantum phase transition by showing that its derivative with respect to anisotropy parameter diverges at the critical point.
We then verify the universality of the critical phenomena of the EPR steerability function in the system.
We also use two-qubit EPR-steering inequality to explore the relation between EPR steering and quantum phase transition.
\end{abstract}


 \maketitle

Einstein {\it et al.} presented the so-called Einstein-Podolsky-Rosen (EPR) paradox to question the completeness of quantum mechanics based on locality and realism in 1935 \cite{EPR}.
Soon after, Schr\"{o}dinger \cite{SEn} introduced the term of entanglement
to describe the correlations between two particles. Entanglement that a quantum state which cannot be separated, is the first type of nonlocal effect identified. In 1964, Bell \cite{Bell} presented a method
in the form of Bell inequality to describe quantum nonlocal property based on
the assumptions of locality and realism. The violation of Bell inequality rules out local
hidden variable theories to describe quantum mechanics as well as implies
the existence of the so-called Bell nonlocality \cite{CHSH,MABK,WWZB,us}, that is the second type of nonlcoal
phenomenon arising from the EPR paradox. For Bell nonlocality, entanglement is necessary but
not sufficient. Recently, EPR steering has been shown to be the third type of nonlocal property \cite{Reid,CV,entropic,NP2010,steering3,steering5,multi}.
%by resorting to EPR-steering inequality \cite{Reid,CV,entropic,NP2010,steering3,steering5,multi}.
EPR steering, like entanglement, was originated from Shr{\"
o}dinger's reply to the EPR paradox to reflect the inconsistency
between quantum mechanics and local realism. For a pure entangled
state held by two separated observers Alice and Bob, Bob's qubit can
be ``steered" into different states although Alice has no access to
the qubit. The EPR steering has been experimentally observed by
violation of EPR-steering inequalities and the violation
demonstrates the impossibility of using local hidden state theories
to complete quantum mechanics \cite{NP2010}. Within the hierarchy of
nonlocality, Bell nonlocality is the strongest, followed by EPR
steering, while entanglement is the weakest
\cite{steering1,steering2}.

In spite of the essential role played by quantum nonlocality in
quantum information and computation, Literatures \cite{EQPT,BQPT}
have shown that both entanglement and Bell nonlocality can be used
as a tool to reveal quantum phase transition (QPT) in many-body
quantum systems. For entanglement, one needs to find out the exact
form of the reduced density matrix of the ground state and utilize
entanglement measure to signal QPTs. While for Bell nonlocality,
which is the most stringent to test experimentally, the reduced
density matrix of the ground state may not violate Bell inequality
and hence one does not have Bell nonlocality, but the Bell function
value still can be used to capture QPTs.
%Another promising approach to capture QPTs by quantum nonlocality is to utilize the EPR steering.
EPR steering lies strictly intermediate between Bell nonlocality and
entanglement \cite{steering1,steering2} and in principle, EPR
steering should be easier observed than Bell nonlocality due to the
asymmetry description between two observers Alice and Bob. It has
been shown experimentally that some Bell-local states exhibit EPR
steering and the EPR-steering inequality allowing for multi-setting
measurements for two parties has been presented \cite{NP2010}.
Although many efforts have been devoted to the investigations of EPR
steering, the EPR-steering inequalities in the literatures are not
strong enough for two-qubit systems. Therefore, it is not possible
to observe the EPR steering for some states, especially for mixed
states. Very recently, a criterion for EPR steering of two qubits
has presented and it has been numerically proved to be a strong
condition to witness steerability \cite{ppt}. This offers an
effective way to detect EPR steering for two qubits. An interesting
question is whether EPR steering can be used to investigate the
behavior of condensed matter systems.

In this work, we investigate the XY spin chain
to establish the relation between EPR steering and QPTs. Utilizing the EPR-steering criterion
proposed in our recent work \cite{ppt}, we find the EPR steerability function $\mathcal{S}$
as defined in Eq. (\ref{steering-criterion}) of reduced density matrix of the ground state. The function $\mathcal{S}$ indicates the existence of EPR steering when it is smaller than $0$.
We analyze the function $\mathcal{S}$ and its nonanalytic behavior at the transition point.
Our results show that the quantum criticality in the XY model can be captured by the EPR steerability function
of the ground state, and this enables conveniently testable quantum nonlocality to signal the QPT.
We also explore the relation between EPR steering and QPT by utilizing two-qubit EPR-steering inequality.

Consider one-dimensional XY spin chain with Hamiltonian given by
\begin{eqnarray}
H=\sum_{-M}^{M}[(1+\gamma)\sigma_i^x\sigma_{i+1}^x+(1-\gamma)\sigma_i^y\sigma_{i+1}^y+h\sigma_i^z],
\end{eqnarray}
where $M=(N-1)/2$ for the spin number $N$ odd, $\gamma$ is anisotropy paramter,
$h$ is the strength of magnetic field, and $\sigma^{x,y,z}_i$ are Pauli operators associated with local spin at site $i$.
The system undergos a QPT at the critical point $h=1$ differing XX-like
phase $\gamma=0$ from Ising-like phase $0<\gamma\leq 1$.

The two-spin reduced
density matrix at sites $i$ and $j$ of the ground state of the spin chain is of the form,
%\begin{eqnarray}
%\rho_{ij}&=&\frac{1}{4}\sum_{a,b=0,x,y,z}\langle \sigma_i^a\sigma_j^b\rangle\sigma_i^a\otimes\sigma_j^b.
%\end{eqnarray}
%where $\sigma^0=\mathcal{I}$ representing two-qubit identity matrix.
\begin{eqnarray}
\rho_{ij}&=&\frac{1}{4}(I+\langle \sigma_i^z\rangle\sigma_i^z\otimes 1
+\langle \sigma_j^z\rangle 1\otimes\sigma_j^z+
\langle \sigma_i^z\sigma_j^z\rangle\sigma_i^z\otimes \sigma_j^z\nonumber \\
&& \sum_{X,Y=x,y}\langle \sigma_i^X\sigma_j^Y\rangle\sigma_i^X\otimes \sigma_j^Y).
\end{eqnarray}
The nonzero correlations $\langle \sigma_{i,j}^z\rangle$ and $\langle \sigma_i^z\sigma_j^z\rangle$ for the XY model are given by \cite{AP16.407}
\begin{eqnarray}
&&\langle \sigma_{i}^z\rangle=-G_0,\nonumber \\
&&\langle \sigma_i^z\sigma_j^z\rangle=G_0^2-G_{i-j}G_{j-i},
\end{eqnarray}
where
\begin{eqnarray}
&&G_r=\frac{1}{M}\sum_1^{M}(-\cos\frac{2\pi k}{N})\cos(r\frac{2\pi k}{N})/\Lambda_{k}\nonumber \\
&&+\frac{\gamma}{M}\sum_1^{M}\sin\frac{2\pi k}{N}\sin(r\frac{2\pi k}{N})/\Lambda_{k},
\end{eqnarray}
with %$M=\frac{N-1}{2}$ and
$\Lambda_{k}=\sqrt{(\gamma\sin\frac{2\pi k}{N})^2+\cos^2\frac{2\pi
k}{N}}$. We need to explore quantum nonanalytic property in
thermodynamic limit when $N\rightarrow \infty$, then the sums in the
expectation values are replaced by integrals,
\begin{eqnarray}
&&G_r=\frac{1}{\pi}\int_0^{\pi}d\phi(-\cos\phi)\cos(\phi r)/\Lambda_{\phi}\nonumber \\
&&+\frac{\gamma}{\pi}\int_0^{\pi}d\phi\sin\phi\sin(\phi r)/\Lambda_{\phi},
\end{eqnarray}
and $\Lambda_{\phi}=\sqrt{(\gamma\sin\phi)^2+\cos^2\phi}$.

The very recent proposed criterion for EPR steering \cite{ppt} is
obtained from the constraints on the eigenvalues of partial
transpose of 2-qubit density operator $\rho_{ij}$. Let
$\{\lambda_1,\lambda_2,\lambda_3,\lambda_4\}$ be four eigenvalues of
$\rho_{ij}^{T_j}$ listed in ascending order, the experienced
condition for $\rho_{ij}$ bearing EPR steering is
\begin{eqnarray}
\mathcal{S}=\lambda_1+\lambda_2-(\lambda_1-\lambda_2)^2<0.\label{steering-criterion}
\end{eqnarray}
If the above eigenvalue combination $\mathcal{S}$ is negative, the EPR steering of $\rho_{AB}$ can be certified. %The function $\mathcal{S}$ can be
%considered as a quantitative measure of EPR steering.
To demonstrate the relation between EPR steering and QPTs, we plot EPR steerability function $\mathcal{S}$ of $\rho_{ij}$ and its derivative with respect
to the field strength $h$ when $\gamma=0.6$ for different values of $N$.
\begin{figure}[tbp]
\includegraphics[width=90mm]{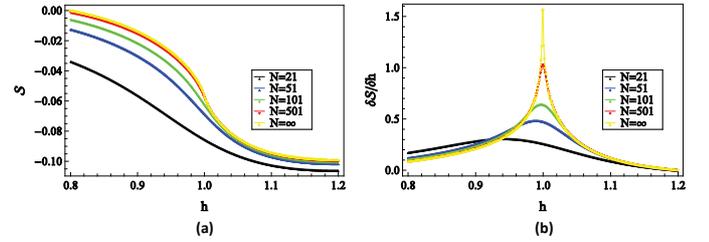}\\
\caption{(Color online) The variance of $\mathcal{S}$ of 2-qubit reduced density matrix $\rho_{ij}$ and its derivative $d\mathcal{S}/dh$ versus $h$ when $\gamma$ is fixed to be $0.6$.
The curves are for different spin sizes $N$}\label{fig1}
\end{figure}
Fig.\ref{fig1} (a) shows the variance of $\mathcal{S}$ with increasing $h$ when $\gamma$ is fixed to be $0.6$.
It is obvious that $\mathcal{S}$ is smaller than $0$ for the range of $h>0.8$, and this tells us that $\rho_{ij}$ exhibits EPR steering.
We plot $d\mathcal{S}/dh$ versus $h$ in Fig.\ref{fig1} (b) for different spin size $N$. The nonanalytic property of the EPR steerability function in the XY model is
clearly shown at the critical point $h_c=1$. Thus, $\mathcal{S}$ of the
ground state is a witness of QPT.

We next explore the scaling behavior of $\mathcal{S}$ by the finite size scaling approach \cite{cri} to further understand the relation between EPR steering and
QPT. %As we know, systems with different microscopic dynamics may behave equivalently at criticality, and the behavior depends only on the dimension of system and the symmetry of order parameter.
In QPT, the critical feature can be characterized by a universal
quantity whose behavior at criticality is entirely
described by a critical exponent $\nu$ in the form of $\xi\sim|\lambda-\lambda_c|^{-\nu}$.
To study quantum criticality in XY model, one needs to
distinguish two universality classes depending on the anisotropy parameter $\gamma$.
For any value of $\gamma$, quantum critical behavior occurs at the transition point $h_c=1$.
For $\gamma=0$ the XY model belongs to XX universality class and critical exponent is $\nu=1/2$, and for $0<\gamma\leq 1$ the model belongs to Ising
universality class and critical exponent is $\nu=1$ \cite{QPT}.
%Suppose that $\mathcal{S}$ is a universal quantity, we should be able to find a
%critical exponent for $\mathcal{S}$. %namely we should be able to write
%\begin{eqnarray}
%\mathcal{S}\propto|h-h_c|^\nu
%\end{eqnarray}

\begin{figure}[tbp]
\includegraphics[width=80mm]{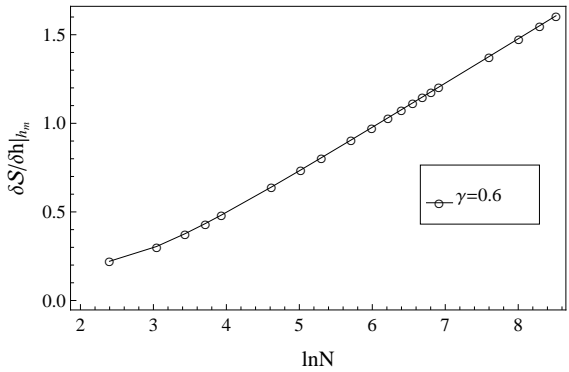}\\
\caption{The maximum value of $d\mathcal{S}/dh$ at the pseudocritical point $h_m$ versus lattice
sizes $\ln N$.}
\label{fig2}
\end{figure}

Let us consider the derivative of steerability function with respect to $h$ as a function $h$ for different spin size $N$ as shown in Fig. \ref{fig1} (b).
For finite spin size, the curves do not have divergence, but show obvious
humps near critical point $h_c=1$. With increasing spin size $N$, the peak of each curve becomes sharper.
Each curve approaches to its maximal value at pseudocritical point $h_m$ which changes towards the critical point $h_c$ when $N$ increases.
In the thermodynamic limit, when $N\rightarrow \infty$, the singular behavior of $d\mathcal{S}/dh$ is clear in the vicinity of the quantum critical point, and it
can be analyzed as,
\begin{eqnarray}
\frac{d\mathcal{S}}{dh}\approx \kappa_1\ln|h-h_c|+{\rm const},
\end{eqnarray}
where $\kappa_1=0.2356$. We then plot the value of $d\mathcal{S}/dh$ at $h_m$ versus spin size $\ln N$ in Fig. \ref{fig2} which shows the relation,
\begin{eqnarray}
\frac{d\mathcal{S}}{dh}|_{h_m}\approx \kappa_2\ln N+{\rm const},
\end{eqnarray}
with $\kappa_2=0.2355$. According to the scaling ansatz in the case
of logarithmic divergence \cite{cri}, the ratio
$|\kappa_1/\kappa_2|$ gives the exponent $\nu$ of $\mathcal{S}$. By
our numerical results, $\nu=1$ is approximately obtained for the XY
model when $\gamma=0.6$.
%\begin{figure}[tbp]
%\includegraphics[width=80mm]{qptt2.eps}\\
%\caption{The maximum value of $d\mathcal{S}/dh$ at the pseudocritical point $h_m$ versus lattice
%sizes $\ln N$. }
%\label{fig3}
%\end{figure}
Therefore, our results show that the EPR steerability function
of the ground state can signal the quantum criticality in the XY model.
%To verify the universality of $\mathcal{S}$ in the XY model, we need to examine the
%scaling behavior for different $\gamma$. We look at the asymptotic behaviors of $d\mathcal{S}/dh$ when $\gamma=0.8$.
%Similarly we find two equations with $\kappa_1=0.1807$ and $\kappa_2=0.1808$ from which we can obtain approximately the same critical exponent
%$\nu=1$.

\begin{figure}[tbp]
\includegraphics[width=90mm]{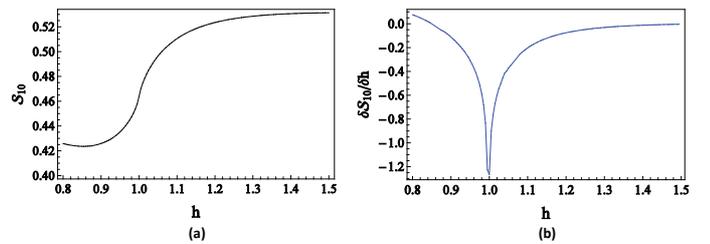}\\
\caption{Quantum predictions of $\mathcal{S}_{10}$ and
$d\mathcal{S}_{10}/dh$ versus $h$ when $\gamma=0.6$ in thermodynamic
limit when $N\rightarrow \infty$.}\label{fig4}
\end{figure}

The known EPR-steering inequalities can also be used to demonstrate
the fact that EPR steering is able to capture quantum criticality in
XY model. Here we consider the $N$-setting EPR-steering inequalities
proposed in Ref. \cite{NP2010} which is based on the assumption that
observer Alice's measurement result is described by the random
variable $A_k=\pm 1\;(k=1,...N)$ and Bob's kth measurement is
defined by Pauli observables $\vec{\sigma}_k^{B}$ along some axis
$\hat{n}_k$, and the two qubit EPR-steering inequality is of the
form
\begin{eqnarray}
\mathcal{S}_N=\frac{1}{N}\sum_{k=1}^N\langle A_k\vec{\sigma}_k^B\rangle\leq C_N,
\end{eqnarray}
%with
%\begin{eqnarray}
%C_N={\rm max}_{\{A_k\}}\{\Lambda_{\rm max}(\frac{1}{N}\sum_{k=1}^N A_k\vec{\sigma}_k^B)\}
%\end{eqnarray}
%where $\Lambda_{\rm max}(O)$ defines the largest eigvenvalue of observable $O$.
where $C_N$ is the limit imposed by local hidden state theoreties.
When $N=2$, $C_2=1/\sqrt{2}$; when $N=3$, $C_3=1/\sqrt{3}$; and when
$N=10$, $C_{10}=0.5236$. It is obvious that the more the number of
measurement settings, the stronger the two-qubit EPR-steering
inequality is. We utilize 10-setting EPR-steering inequality to
investigate the EPR steering of XY spin chain and plot quantum
prediction of $\mathcal{S}_{10}$ and its derivative with respect to
$h$ in thermodynamic limit when $N\rightarrow \infty$ in Fig.
\ref{fig4}. From Fig. \ref{fig4} (a), we find that for some values
of $h$, the quantum predictions do not exceed $C_{10}$ and the
10-setting EPR-steering inequality cannot identify the EPR steering
of the ground state in the vicinity of critical point. Even if no
violation is found, the derivative of quantum prediction of
$\mathcal{S}_{10}$ still exhibits singular property at the critical
point $h_c$, as shown in Fig. \ref{fig4} (b). In a word,
$\mathcal{S}_{10}$ is able to signal the nonanalytic features in the
XY spin chain when quantum prediction of $\mathcal{S}_{10}$ is
smaller than $C_{10}$, or no EPR steering identified by the
inequality. The result is similar to that for Bell's inequality in
QPT \cite{BQPT} that Bell function value can capture QPTs although
Bell's inequality is not violated. On the other hand, according to
the EPR steering criterion $\mathcal{S}$, it is found that the
density matrix of the ground state has EPR steering. This suggests
that the 10-setting EPR-steering inequalities are not strong enough
to detect EPR steering in the XY model and so it is not a tight
inequality for EPR steering.
%EPR-steering inequalities are a subset of entanglement witnesses and a superset of
%Bell's inequalities. Bell's inequalities are the most stringent constraint among the three ones, and in principle EPR-steering
%inequalities are easier to be violated experimentally than Bell's inequalities.
We expect more effective EPR-steering inequalities which can enable to detect EPR steering in QPT, and this will make it convenient to demonstrate experimentally
the connection between EPR steering and QPT.

%\begin{eqnarray}
%S_N=\frac{1}{N}\sum_{k=1}^N\langle A_kB_k\vec{\sigma}_k^C\rangle\leq \mathcal{C}_N
%\end{eqnarray}
%with
%\begin{eqnarray}
%C_N={\rm max}_{\{A_k\}}\{\Lambda_{\rm max}(\frac{1}{N}\sum_{k=1}^N A_k\vec{\sigma}_k^B)\}
%\end{eqnarray}

\vspace{3mm} To summarize, we have investigated the relation between
EPR steering and QPT in the anisotropic spin-1/2 XY model by using
the 2-qubit EPR steering criterion. The EPR steerability function
$\mathcal{S}$ shows the existence of the EPR steering of the ground
state of the model. As the spin number goes to infinity, the system
undergoes QPT between the spin-fluid and the Ising-like phases,
which can be captured by $\mathcal{S}$. By studying the nonanalytic
behavior of $\mathcal{S}$ in the vicinity of transition point
$h_c=1$, we find that $\mathcal{S}$ is a universal quantity to
describe QPT in the XY model, and this makes it possible to
demonstrate experimentally the connection between EPR steering and
QPT. The result that EPR steerability function is able to signal QPT
can be understood as follows. The function $\mathcal{S}$ is the
combination of eigenvalues of partial transpose of $\rho_{ij}$ which
changes dramatically at the transition point, and the information of
the critical change is obviously contained in the eigenvalues of
$\rho_{ij}$ or its partial transpose. Thus, the EPR steerability
function $\mathcal{S}$ can reflect the critical feature in QPT. We
believe that the result is applicable to other quantum many-body
systems. We also discuss the relation between EPR steering and QPT
by resorting to 10-setting EPR-steering inequality. Although the
EPR-steering inequality is not violated near the critical point,
quantum prediction of left-hand-side of the inequality still
exhibits singular behavior. This suggests two particular points: (1)
Quantum prediction of left-hand-side of EPR-steering inequalities
plays a interesting role to capture QPT no matter whether the
inequalities are violated or not, just like the case for Bell's
inequalities in QPT \cite{BQPT}; (2) The present two-qubit
EPR-steering inequalities are not strong enough to detect EPR
steering in the XY model. Our results from EPR steering criterion
show that the reduced density matrix of the ground state of XY spin
chain bears EPR steering and it indeed signals the QPT. We expect
more effective EPR-steering inequalities which can enable to detect
EPR steering in QPT.

\vspace{3mm}

%\centerline{\textbf{ACKNOWLEDGEMENTS}}

J.L.C. is supported by National Basic Research Program (973 Program)
of China under Grant No. 2012CB921900 and NSF of China (Grant Nos.
10975075 and 11175089). This work is also partly supported by
National Research Foundation and Ministry of Education, Singapore
(Grant No. WBS: R-710-000-008-271).

\end{document}